\documentclass{article}    

\usepackage{graphicx}
\usepackage[utf8]{inputenc}
\usepackage{amsmath, amssymb} 
\usepackage{mathtools} \usepackage{color}
\usepackage[left,modulo]{lineno}
\usepackage{algorithm}
\usepackage{algorithmic}
\usepackage{nicefrac}
\usepackage{dsfont}
\usepackage[round, sort]{natbib}
\usepackage{hyperref}
\usepackage{nicematrix}
\usepackage{soul}

\newcommand \OU {Ornstein-Uhlenbeck }
\newcommand \ou [1]{{#1}_{\text{ou}}}
\newcommand \oui [1]{{#1}_{\text{ou},i}}
\newcommand \RR {\mathbb{R}}
\newcommand \dx [2][]{\mathrm{d}^{#1}\mspace{-1mu}\mathord{#2}}

\makeatletter
\newcommand \unif {\@ifstar {\@unifstarred} {\@unifnostar}}
\newcommand \@unifstarred {\mathcal{U}}
\newcommand \@unifnostar [1]{\unif* \left( {#1} \right)}
\makeatother
\makeatletter
\newcommand \normal {\@ifstar {\@normalstarred} {\@normalnostar}}
\newcommand \@normalstarred [1][]{\mathcal{N}_{{#1}}}
\newcommand \@normalnostar [3][]{\normal*[{#1}] \left({#2}, {#3}\right)}
\makeatother
\newcommand \pv {\mathfrak{p}}
\newcommand \qv {\mathfrak{q}}
\newcommand \zs {\mathfrak{z}}
\newcommand \ts {\mathfrak{t}}

\newcommand \tree {\mathcal{T}}
\newcommand \tips {\{T_1, \dots, T_{m}\}}
\newcommand \shifts {\Delta}
\newcommand \optim {\beta}

\newcommand \nodes {\{N_1, \dots, N_{n-m}, T_1, \dots, T_{m}\}}
\newcommand \shiftset {\mathcal{D}}
\DeclareMathOperator*{\argmin}{argmin}

\begin{document}

\title{Hierarchical correction of p-values via an ultrametric tree running
  Ornstein-Uhlenbeck process}

\author{Antoine Bichat\,$^{1,2}$,\\ Christophe Ambroise\,$^{1}$ and Mahendra Mariadassou\,$^{3,*}$\\
$^{1}$LaMME, Université d’Évry val d’Essonne, 91000 Évry, France   \\
$^{2}$Enterome, 94-96 Avenue Ledru Rollin, 75011 Paris, France\\
$^{3}$MaIAGE, INRAE, Université Paris-Saclay, 78350, Jouy-en-Josas, France}

\setlength{\parindent}{0ex}

\maketitle

\begin{abstract} Statistical testing is classically used as an exploratory tool to search for
  association between a phenotype and many possible
  explanatory variables. This approach often leads to multiple  testing
  under dependence.

We assume a hierarchical structure between tests via an \OU process on a tree.  The process correlation structure is used for smoothing the $p$-values. We  design  a  penalized estimation of the mean of the \OU process for $p$-value computation.

The performances of the algorithm are assessed via
  simulations. Its ability to discover new associations is demonstrated on a metagenomic dataset.

The corresponding R package is available from \url{https://github.com/abichat/zazou}.

\end{abstract}

\section{Introduction}
\label{sec:Introduction}

In many fields, statistical testing is classically used as an
exploratory tool to look for the association between a variable of interest
 and many possible explanatory variables. For example, in
transcriptomics, the link between a phenotype and the expression of
tens of thousands of genes is tested
\citep{mclachlan2005analyzing}, in Genome Wide Association
Studies (GWAS) the association between millions of markers and a
phenotype is tested \citep{bush2012genome}, in functional Magnetic Resonance Imaging (fMRI),
the goal is to identify voxels that are significantly activated in
two different conditions \citep{cremers2017relation}.

This problem of multiple comparisons dates back to the work of Tukey
 \citep{tukey1953problem}.  It has since been the subject of abundant
 literature and aims at controlling a probability of error of some
 sort. Most of the literature focus on the control of the Familiy
 Wise Error Rate (FWER) \citep{bland1995multiple}, being the
 probability of at least one false discovery among detections, or of the False
 Discovery Rate (FDR) \citep{benjamini1995controlling}, defined as the
 expected proportion of false positives among detections. 

Most of the correction procedures for controlling FWER or FDR, such as the popular Benjamini-Hochberg (BH) procedure, rely on
independence, or some form of weak dependence, among the hypotheses, which
is rarely observed in practice.
Multiple testing under dependence is a
difficult problem occurring in many fields. In transcriptomics,
differential analysis has to deal with gene expressions that are often
highly correlated. When performing GWAS, the linkage desiquilibrium
imposes a strong spatial dependence between markers, and in Functional
Magnetic Resonance Imaging (fMRI), two spatially close voxels have
often comparable activation.

The control of the FDR remains valid under arbitrary dependency structures by replacing the BH procedure with the more conservative BY procedure of 
\citet{benjamini2001control}. However, based on results obtained from simulated datasets, it is obvious that
there is a substantial loss of power when the real dependency structure is not taken into account, as discussed in depth in \citet{blanchard2020dependency}.

An alternative approach  for dealing with multiple testing  is to reduce the number of tests by
aggregating certain hypotheses. Aggregation strategies 
vary and can be based on a priori knowledge (\emph{e.g.} metabolic pathways, functional modules of genes) or on clustering algorithms \citep{sankaran2014structssi, renaux2020hierarchical}.

This article aims to take into account the dependencies between variables in order to offer a powerful statistical procedure of multiple testing. A hierarchical dependency structure between variables is assumed to be known up to certain constants. This assumption is common in our motivating example of microbiome studies \citep{sankaran2014structssi, xiao2017false, huang2021treeclimbr,matsen2013edgepca,silverman2017philr}, where the phylogeny is a natural hierarchical structure encoding similarities between variables (or namely species in that context).
The hypotheses tested can then be organized in a tree structure which captures correlations at different scales of observation. This type of hierarchical structure is observable in transcriptomics differential analysis, where gene expressions can easily be
represented by a hierarchy based on gene expression correlation. In GWAS and fMRI, spatial dependence also proves to be very suitable for hierarchical modeling \citep{ambroise2019adjacency,eickhoff2015connectivity, sesia2020multi}.

We propose to model the hierarchical structure of the multiple tests through an \OU process on a tree. The process correlation structure is used for smoothing the $p$-values, after conversion to $z$-scores, similarly to the algorithm proposed in \citet{xiao2017false} but with an explicit underlying model. 

We then consider a three stage approach for our differential analysis procedure. 
The first stage reframes the initial problem as a linear regression problem that preserves the hierarchical structure. This linear problem is ill defined ($p \sim 2n$) and we therefore resort to an $\ell_1$ penalized estimation of the mean of the \OU process. 
The second stage produces asymptotically valid $p$-values. The output of $\ell_1$ penalized estimation produces are indeed biased and offer no theoretical guarantees about their asymptotic distribution; we therefore correct them using a debiasing procedure \citep{javanmard2013confidence, javanmard2014confidence,
zhang2014confidence} to compute valid $p$-values. The third and final stage controls the FDR of the overall procedure, using the tuning strategy of  \citet{javanmard2019false}.

The selection strength of the \OU process and the penalty parameter are hyperparameters  of our model, whose selection  is achieved via a Bayesian Information Criterion (BIC). We provide some background on hierarchical procedures in Section~\ref{sec:Background}, introduce the model and statistical procedure in Section \ref{sec:Model} and detail the computational steps in Section~\ref{sec:Optimization}. The performances of the
algorithm are assessed via simulations in Section \ref{sec:Simul}. The
use of the proposed model is illustrated in Section \ref{sec:Appli},
where we demonstrate its ability to discover novel associations in a metagenomic dataset.
 
\section{Background}
\label{sec:Background}

\subsection{Examples of multiple testing strategies}

A classic example in genomics consists in grouping the markers
according to whether they belong to the same genes (aggregation by a
prior). The genes can then be grouped according to their similarity, computed for example from expression profiles. \citet{kim2010spatial} have, for example,
proposed a hierarchical testing strategy controlling the FWER in a hierarchical manner, by testing clusters of genes, then individual genes associated with a phenotype with the goal of finding genomic regions associated with  a specific type of cancer. This type of top-down
approach uses the concept of sequential rejection principle \citep{goeman2012inheritance, meinshausen2008hierarchical,renaux2020hierarchical}.

fMRI is another domain where tests are aggregated: neighboring voxels that are highly correlated are aggregated into a single voxel cluster. \citet{benjamini2007false} propose an adaptation of the False
Discovery Rate (FDR) to allow for cluster-level multiple testing for fMRI data.

\emph{Ad hoc} aggregating  methods for multiple testing also exist in Metagenomics.  LEfSe \citep{segata2011metagenomic} performs a bottom
up approach where a factorial Kruskal-Wallis rank sum test is applied to each feature with respect to a class factor,  followed by a pairwise Wilcoxon test, and a linear discriminant analysis.  MiLineage \citep{tang2017general} performs multivariate tests concerning multiple taxa in a lineage to test the association of lineages to a phenotypic outcome.

\subsection{Independence assumption}

The assumption of independence of tests is convenient as it enables for both exact analyses and simple error bounds for classical procedures \citep[e.g.]{benjamini1995controlling}. It is however unrealistic in practice. In many fields, including all the previous examples, measurements typically exhibit strong correlations.  
Some correction procedures, like the one proposed by
\citet{benjamini2001control}, make few assumptions while  guaranteeing control of the FDR. Those general guarantees come with a high cost in terms of statistical power: the nominal FDR typically is much smaller that the target, resulting in many FN. Permutation procedures are an appealing alernative that can automatically adapt to the dependence structure of the p-values \citep{tusher2001significance} but may fail when confronted to unbalanced design or correlated data. Knowledge of the correlation structure
can be leveraged to increase the power while still controling the FDR below a given target. Several approaches have been developed along those
lines when the tests are organized along a hierarchical structure, typically encoded in a tree.

\subsection{Hierarchical testing}

The Hierarchical FDR (hFDR) introduced by   \citet{yekutieli2008hierarchical} and implemented in the R package
\texttt{structSSI} \citep{sankaran2014structssi}, proposes a top-down
algorithm to sequentially reject hypotheses organized in a tree. The same approach is used in \citep{renaux2020hierarchical} to select a group of variables arranged in a clustering tree. However,
this approach suffers from some limitations, as shown in \citep{bichat2020incorporating, huang2021treeclimbr}. First, the algorithm in its vanilla formulation commonly fails to move down on the tree because of failure to reject the topmost node. Second, it only controls for
an \textit{a posteriori} FDR level, which is a complex function of the (user-defined) \textit{a priori} FDR
level and the structure of rejected nodes. This makes it difficult to calibrate the \textit{a priori} FDR that would achieve a target \textit{a posteriori} FDR and thus to compare it to other correction methods. Finally, it does not produce a corrected $p$-value, or $q$-value, per leaf, but only a \emph{reject / no reject} decision and was shown in \citep{bichat2020incorporating} to perform no better than BH in many instances. Given all these drawbacks, we did not include the hFDR in our benchmark and use BH as a baseline instead.

\texttt{StructFDR} \citep{xiao2017false} was developed for metagenomics Differential Abundance Testing (DAT) and 
relies on $z$-scores / $p$-values smoothing followed by permutation correction. Given any taxa-wise DAT procedure, $p$-values $\pv$ are first computed for all $m$ taxa (\emph{i.e.} leaves of the tree) and then transformed to $z$-scores $\zs$. The tree is used  to compute a distance matrix $\left(\mathbf{D}_{i,j}\right)$ and then turned into a correlation matrix $\mathbf{C}_{\rho} = \left(\exp\left(-2\rho \mathbf{D}_{i,j}\right)\right)$ between taxa using a Gaussian kernel. The $z$-scores are then smoothed using the following hierarchical model:
$$\left. \zs \mid \mu \right. \sim \mathcal{N}_m\left(\mu,\sigma^2\mathbf{I}_m\right),$$
$$\mu \sim \mathcal{N}_m\left(\gamma \mathbf{1}_m,\tau^2\mathbf{C}_{\rho}\right),$$
where $\mu$ captures the effect size of each taxa and $\zs$ is a noisy observation of $\mu$. The maximum a posteriori estimator $\mu^*$ of $\mu$ is given by
\begin{equation*}\mu^* = \left(\mathbf{I}_m+ k\mathbf{C}_{\rho}^{-1}\right)^{-1}\left(k\mathbf{C}_{\rho}^{-1}\gamma\mathbf{1}_m+\zs\right) \quad \text{where} \quad k = \sigma^2 / \tau^2 .
\end{equation*}
The FDR is controlled by means of a resampling procedure to estimate the distribution of $\mu^*$ under $H_0$ and estimate adjusted $p$-values $\qv^{\text{sf}}$. This method is implemented in the \texttt{StructFDR} package \citep{structfdr2018}.

\texttt{TreeclimbR} \citep{huang2021treeclimbr}
is a bottom-up approach also developed for metagenomics DAT but with a broader scope. It relies on aggregating abundances at each node of the tree (understood as a cluster of taxa) and performing a test to compute one $p$-value per node (compared one test per leaf for \texttt{StructFDR}). The main idea is then to use those $p$-values to compute a score for node $i$
$$
U_i(t) =
\left|\frac{\sum_{k\in B(i)} \mathfrak{s}_k \mathds{1}_{\{\pv_k \leq t\}}}{\#B(i)}\right|
$$
where $B(i)$ is the set of descendants of node $i$, $\pv_k$ and $\mathfrak{s}_k \in \{-1, +1\}$ are the p-value of the node $k$ and the sign of the associated effect, and $t$ is a tuning parameter. A node $i$ will be considered as candidate if $U_i(t) \simeq 1$ and $\pv_i < \alpha$. This ensure that all descendants are (i) significant at level $t$ with (ii) effects of coherent sign. At the end, multiplicity correction is only done on nodes (including leaves) that do not descend from another candidate.

\section{Models and algorithms \label{sec:Model}}

Our correction methods assumes that $p$-values, or rather z-scores, evolve according to an \OU process on a tree. We thus use the corresponding correlation structure to decorrelate the $z$-scores and, in turn, the $p$-values. This is similar in spirit to the smoothing algorithm of \cite{xiao2017false} but we derive our procedure from first principles and explicit assumptions. We first remind a few properties of \OU processes before proceeding to   our model and procedure. 

\subsection{Ornstein-Uhlenbeck process on a tree}

An \OU (OU) process $(W_t)$ with optimal value (also called drift) $\ou{\optim}$, selection strengh (also called mean reversion parameter) $\ou{\alpha}$ and variance of the white noise $\ou{\sigma^2}$, is a Gaussian process that satisfies the stochastic differential equation: 
\begin{equation*}
\dx{W_t} = - \ou{\alpha} (W_t -  \ou{\optim}) \dx{t} + \ou{\sigma}\dx{B_t}.
\end{equation*}

\begin{figure}[h!]
\begin{center}
\includegraphics[width=\linewidth]{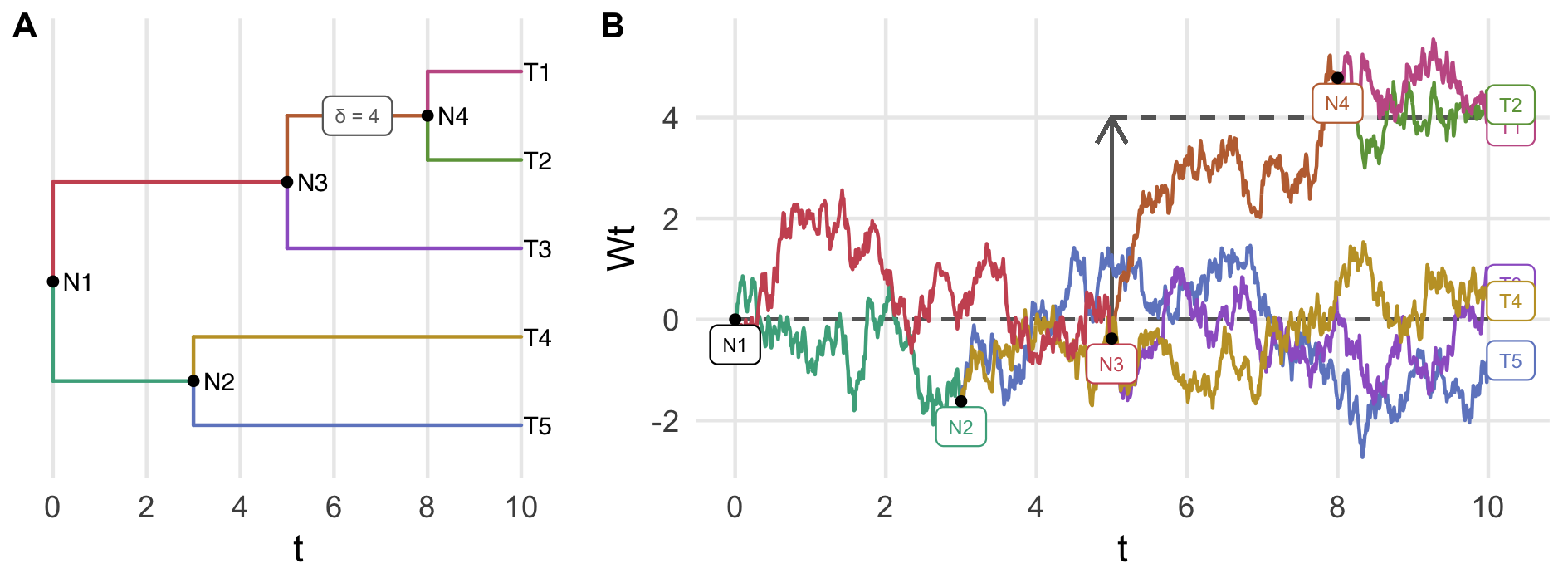}
\end{center}
\caption{\label{fig:treeou}\textbf{(A)} Phylogenetic tree with 5 leaves and 4 internal nodes (root $N_1$ included). A shift occurs on the branch leading to $N_4$. \textbf{(B)} \OU process with shifts on the tree defined in the left panel. At each node, the process spawns two independent process with the same initial value. The shifts on the optimal value on the branch leading to $N_4$ results in a different mean value for $N_4$ and all its offsprings ($T_1$ and $T_2$).}
\end{figure}

The important properties of OU processes are bounded variance and convergence to a stationary distribution centered on the optimal value $\ou{\optim}$, namely $W_t\xrightarrow[]{(d)} \normal{\ou{\optim}}{\ou{\sigma}^2/ 2\ou{\alpha}}$ when $t \to \infty$. Thanks to those properties, OU processes have become a popular model applied in various subfields of biology, ranging from evolution of continuous traits, such as body mass \citep{Freckleton2003}, fitness \citep{Lande1976} or CpG enrichment in viral sequences \citep{maclean2021natural} to animal movement \citep{Dunn1977} and epidemiology \citep{Nasell1999}. They naturally emerge as the continuous limit of broad range of discrete-time  evolution models \citep{Lande1976}. \OU processes can be readily adapted to tree-like structures as illustrated in Fig.~\ref{fig:treeou}. 

Formally, we consider a rooted ultrametric tree $\mathcal{T}$ with $m$ leaves and $n$ branches ($n = 2m - 1$ for binary trees). The internal nodes are labeled $N_1$ (the root) to $N_{n-m}$ and the leaves $T_1$ to $T_m$. Let $i$ be a node, $W_i$ the value of the trait at that node and denote $pa(i)$ its unique parent. By convention, we set $t_{N_1} = 0$ and assume $W_{N_1} = 0$. The branch leading to $i$ from $pa(i)$ is denoted $b_i$ and has length $l_i = t_i -t_{pa(i)}$ where $t_i$ is the time elapsed between the root and node $i$. Since the tree is ultrametric, $t_i = h$ for all $i \in \tips$. For any pair of nodes $(i,j)$, let $t_{ij}$ be the time elapsed between the root and the most recent common ancestor of $i$ and $j$ and denote $d_{ij} = t_i - t_j -2t_{ij}$ the distance in the tree between nodes $i$ and $j$. The distribution of the trait at node $i$ is given by:

\begin{equation}
\label{eq:ou_equation}
 W_{i} | W_{pa(i)} \sim \mathcal{N} \left(\lambda_i W_{pa(i)} + (1 - \lambda_i) \oui{\optim}, \frac{\ou{\sigma}^2}{2\ou{\alpha}} (1 - \lambda_i^2) \right) 
\end{equation}
where $\lambda_i = \exp(-\ou{\alpha} l_i)$ and $\oui{\optim}$ is the optimal value on branch $i$. Remark that the process mean value does not immediately shift to $\oui{\optim}$ but lags behind it with a shrinkage parameter controlled by $1 - \lambda_i$. If $\oui{\optim} = 0$ for all $i$, straightforward computations show that $W = (W_{T_1}, \dots, W_{T_m})$ is a gaussian vector with distribution 
\begin{equation*}
W \sim \mathcal{N}(0, \Sigma) \quad \text{where} \quad \Sigma_{ij} = \frac{\ou{\sigma}^2}{2\ou{\alpha}} e^{-2\ou{\alpha}d_{ij}} (1 -  e^{-2\ou{\alpha}t_{ij}}).
\end{equation*}

When, the optimal value can shift on a branch (\emph{e.g.} the branch $b_{N_4}$ leading to $N_4$ in Fig.~\ref{fig:treeou}), the mean vector of $W$ is a slightly more complex and depends on both the tree topology and the location and magnitude of the shifts. Denote $U$ the $m \times (n+m)$ incidence matrix of $\mathcal{T}$ with rows labeled by leaves ($i \in \tips$) and columns labeled by inner nodes and leaves ($j \in \nodes$), with entries defined as $U_{ij} = 1$ if and only if leaf $i$ is in the subtree rooted at node $j$. Intuitively, column $U_{.j}$ encodes all leaves descending from node $j$ and row $U_{i.}$ encodes all ancestors of leaf $i$. Denote $\shifts$ the dimension $n$ column vector with entries defined as $\shifts_{i} = \oui{\optim} - \optim_{\text{ou},pa(i)}$ where $i \in \nodes$. Non-zero entries of $\shifts$ correspond to \emph{shifts location}, nodes for which the optimal value $\oui{\optim}$ differ from its parent's and their values to \emph{shifts magnitude} (see Figure~\ref{fig:incidence} for an example). Finally let $\Lambda$ be the $n$ diagonal matrix with diagonal entries $\Lambda_{i} = 1 - \exp(\ou{\alpha}(h - t_{pa(i)}))$ where $i \in \nodes$. Straightforward computations (see \cite{bastide2017detection} for detailed derivations) show that $W$ is a gaussian vector with joint distribution:
\begin{equation}
W \sim \mathcal{N}(\mu, \Sigma) \quad \text{where} \quad \mu = U \Lambda \shifts \quad \text{and} \quad\Sigma_{ij} = \frac{\ou{\sigma}^2}{2\ou{\alpha}} e^{-2\ou{\alpha}d_{ij}} (1 -  e^{-2\ou{\alpha}t_{ij}}).
\end{equation}

\begin{figure}
\[
U = 
    \bordermatrix{
    \,\,&N_1\!\!\!&N_2\!\!\!&N_3\!\!\!&N_4\!\!\!&T_1\!\!\!&T_2\!\!\!&T_3\!\!\!&T_4\!\!\!&T_5\cr
    T_1\,\,&1&0&1&1&1&0&0&0&0\cr
    T_2\,\,&1&0&1&1&0&1&0&0&0\cr
    T_3\,\,&1&0&1&0&0&0&1&0&0\cr
    T_4\,\,&1&1&0&0&0&0&0&1&0\cr
    T_5\,\,&1&1&0&0&0&0&0&0&1\cr
    }
    \qquad
    \Delta = \bordermatrix{
    &\cr
    b_{N_1}&0\cr
    b_{N_2}&0\cr
    b_{N_3}&0\cr
    b_{N_4}&\delta\cr
    b_{T_1}&0\cr
    b_{T_2}&0\cr
    b_{T_3}&0\cr
    b_{T_4}&0\cr
    b_{T_5}&0\cr
    }
    \qquad 
    \mu = \bordermatrix{
    &\cr
    \mu_{T_1}&\delta \Lambda_{N_4}\cr
    \mu_{T_2}&\delta \Lambda_{N_4}\cr
    \mu_{T_3}&0\cr
    \mu_{T_4}&0\cr
    \mu_{T_5}&0\cr
    }
\]
\caption{\label{fig:incidence} Incidence matrix $U$, shift vector $\Delta$ and mean vector $\mu$ associated with Fig.~\ref{fig:treeou}. $\Lambda_{N_4} = 1 - e^{\ou{\alpha}(h - t_{N_3})}$ is the shrinkage parameter from equation~\eqref{eq:ou_equation}.}
\end{figure}

When $\mathcal{T}$ is known, the matrix $T = U \Lambda$ is completely specified up to parameter $\ou{\alpha}$. The shifted \OU model, with parameters $\ou{\alpha}$, $\ou{\sigma}^2$ and shift vector $\shifts$, has been used \citep{bastide2017detection, khabbazian2016fast} to find adaptive events, modeled as non zero values in $\shifts$, in the evolution of continuous traits of interest (turtle shell size, great monkey brain shape, etc). In this work, we apply the same mathematical framework to the joint distribution of $p$-values transformed to $z$-scores.

\subsection{Procedure}

We show here how to use the previously described \OU process to incorporate the tree structure $\mathcal{T}$ in the correction of the $p$-values vector $\pv$. 

\paragraph*{Framework.}

Noting $m_{i}^1$ (resp. $m_{i}^2$) the median count (or relative abundance) of taxon $i$ under condition $1$ (resp. condition 2), we want to test $\mathcal{H}_{i0}: m_{i}^1 = m_{i}^2$ against $\mathcal{H}_{i1}: m_{i}^1 \neq m_{i}^2$ and assume that we have a testing procedure that outputs $p$-values, e.g. the Wilcoxon-Mann-Whitney test \citep{mann1947test,wilcoxon1992individual}. We first convert the $p$-values to $z$-scores using the quantile function $\Phi^{-1}$ of the standard gaussian:

$$\zs= \Phi^{-1}(\pv).$$ 

Provided the use of a correct statistical test, we known that $\pv_i \sim \mathcal{U}([0, 1])$ under $\mathcal{H}_{i0}$, so that  $\zs_i \sim \mathcal{N}(0, 1)$. We also know that $\pv_i \preccurlyeq \mathcal{U}([0, 1])$ and thus $\zs_i \preccurlyeq \mathcal{N}(0, 1)$ under $\mathcal{H}_{i1}$. We could also test $\mathcal{H}_{i0}: m_{i}^1 = m_{i}^2$ against $\mathcal{H}_{i1}: m_{i}^1 < m_{i}^2$ or $\mathcal{H}_{i1}: m_{i}^1 > m_{i}^2$, we only require the procedure to output $p$-values that satisfy the previous distributional assumptions for these $\mathcal{H}_{i0}$ and $\mathcal{H}_{i1}$. Note that, even if the test statistic is itself a $z$-score before being transformed to a $p$-value, the $z$-score $\zs_i$ may differ from the \emph{raw} test statistic $z_i$ because of the intermediate $p$-value $\pv_i$. Indeed when considering the simple case of testing equality of means in two samples of size $n$, with gaussian distributions and known variance $\sigma$, the relation between $\zs_i$ and $z_i = \sqrt{n}(\hat{m}_i^1 - \hat{m}_i^2)/2\sigma$ is given by:

$$
\zs_i = \Phi^{-1}(\pv_i) = 
\begin{cases}
\Phi^{-1}(\Phi(z_i)) = z_i & \text{if } \mathcal{H}_{i1}: m_{i}^1 < m_{i}^2 \\
\Phi^{-1}(\Phi(1 - z_i)) = -z_i & \text{if } \mathcal{H}_{i1}: m_{i}^1 > m_{i}^2 \\
\Phi^{-1}(2\Phi(- |z_i|)) & \text{if } \mathcal{H}_{i1}: m_{i}^1 \neq m_{i}^2 \\
\end{cases}
$$ 
After transformation, the test can be thus always be reframed as one-sided on $\zs_i$: $\mathcal{H}_{i0}: E[\zs_i] = 0$ against $\mathcal{H}_{i1}: E[\zs_i] < 0$. We make two assumptions regarding the distribution of $\zs$. 

\begin{itemize}
 \item[(A1)] Under $\mathcal{H}_{i1}$, $\zs_i \sim \mathcal{N}(\mu_i, 1)$ where $\mu_i \leq 0$;
 \item[(A2)] $\zs$ arises from a shifted \OU process on an ultrametric tree $\mathcal{T}$ with parameters $\ou{\alpha}$, $\ou{\shifts}$ and $\shifts$. 
\end{itemize}

Assumption (A1) is very classic when working with $z$-scores \citep{mclachlan2000finite}: finding the alternative hypotheses is equivalent to finding the negative entries of $\mu$. Assumption (A2) allows us to specify the joint distribution of $\zs$ as:
\begin{equation}
\zs \sim \normal[m]{\mu}{\Sigma} 
\end{equation}
where $\Sigma$ is fully specified by the parameters $\ou{\sigma}$ and $\ou{\alpha}$. Note that the diagonal coefficients of $\Sigma$ are all equal to $\ou{\sigma}^2 / 2\ou{\alpha} (1 - 2e^{-2\ou{\alpha}h})$. As they correspond to marginal variances, this forces the equality $\ou{\sigma}^2 =  (1 - 2e^{-2\ou{\alpha}h}) / 2\ou{\alpha}$ so that $\Sigma$ depends only on $\ou{\alpha}$, \emph{i.e.} $\Sigma = \Sigma(\ou{\alpha})$. Finally, the decompositon $\mu = T \shifts$, where $T$ acts as a phylogenetic design matrix, ensures that alternative hypotheses are likely to form clades, \emph{i.e.} groups of leaves obtained by cutting a single branch in the tree. 

This framework allows us to use $\tree$ as a prior structure in the mean vector $\mu$ and variance matrix $\Sigma$ and to recast the hypothesis testing problem as a regression problem. 

\subsubsection{Parameter Estimation}

\paragraph*{Estimation of $\hat{\mu}$. } 

Assume first that $\Sigma$, or equivalently $\ou{\alpha}$, is known. Our main goal is to estimate the negative components of $\mu$. 

To leverage the known tree structure, we use the decomposition $\mu = T\shifts$ and estimate $\mu$ by means of $\shifts$. Since $\shifts$ has dimension $n$ compared to dimension $m$ for $\mu$, we force $\hat{\shifts}$ to be sparse using a constrained lasso penalty \citep{tibshirani1996regression} :
\begin{equation}
\label{eq:plasso}
\hat{\shifts} = \argmin_{\shifts\in \RR^{n} \; \text{s.t.} \; T\shifts \in\RR^m_-} \frac{1}{2} \left\|\zs - T\shifts\right\|_{\Sigma^{-1},2}^2 + \lambda \|\shifts\|_1.
\end{equation}
where $\RR_- = \{x \in \RR \; \text{s.t.} \; x \leq 0\}$.

Intuitively, the decomposition together with the $\ell_1$ penalty works as a nested group lasso penalty for the components of $\mu$, where the groups  correspond to clades of $\tree$, while the constraint $T\shifts \in\RR^m_-$ forces components of $\mu$ to be non positive. For compacity, we define the feasible set $\shiftset = \{ \shifts \in \RR^n \; \text{s.t.} \; T\shifts \in \RR_-^m\}$. Finally, we use the Cholesky decomposition $\Sigma^{-1} = R^TR$ to  simplify the problem into the very well studied optimisation problem:

\begin{equation}
\label{eq:lasso}
\hat{\shifts} = \argmin_{\shifts \in \shiftset} \frac{1}{2} \left\|y - X\shifts\right\|_2^2 + \lambda \|\shifts\|_1
\end{equation}

with $y = R\zs \in \RR^{m}$ and $X = RT \in \RR^{m \times n}$. Note that $y$ is a whitened version of $\zs$, with independent components and spherical covariance matrix. This is a lasso problem with a convex feasability constraint on $\shifts$. The optimisation algorithm used to solve this problem is detailed in Section~\ref{sec:Optimization}.

\paragraph*{Estimation of $\hat{\Sigma}$ and tuning of $\lambda$. }

Remember first that $\Sigma$ is completely determined by $\ou{\alpha}$ because of the link between $\ou{\alpha}$ and $\ou{\sigma}^2$. There are no closed-form expression for the maximum likelihood estimator of $\ou{\alpha}$. We therefore resort to numerical optimisation. To tune the parameter $\lambda$, we test several values to estimate models with different sparsity levels and select the best one using a modified BIC criterion: 

\begin{equation}
 \label{eq:bic}
 (\ou{\hat{\alpha}}, \hat{\lambda}) = \argmin_{\alpha > 0, \lambda \geq 0} \left\|\zs - T\shifts_{\alpha, \lambda}\right\|_{\Sigma^{-1}(\alpha),2}^2 + \log|\Sigma(\alpha)| + \|\shifts_{\alpha, \lambda}\|_0 \log(\log{m})\log{m}
\end{equation}

where $\shifts_{\alpha, \lambda}$ is the solution of problem~\eqref{eq:plasso} for $\Sigma(\alpha)$ and $\lambda$. In practice, $\alpha$ and $\lambda$ vary in a bidimensional grid and we select the values that minimize the objective. We use a modified BIC, where $\log(\log{m})\log{m}$ replaces $\log{m}$, to account for the fact that $m$ scales like $n$ as suggested in \citet{fan2013}.
\newline

\subsubsection{Confidence intervals}

Lasso procedures are known to produce biased estimators and do not return confidence intervals for the point estimate $\hat{\mu}_i$. Instead of simply returning all negative components of $\hat{\mu} = T\hat{\shifts}$, we first debias the estimates and construct confidence intervals for the components of $\shifts$, and in turn of $\hat{\mu}$, using the debiasing procedure of \citet{javanmard2013confidence, javanmard2014confidence, zhang2014confidence}. 

\paragraph*{Debiasing. }

All debiasing procedures assume a model $Y \sim \normal[m]{X\shifts}{\sigma^2 I_m}$ and require both an initial estimator $\hat{\shifts}^{\text{(init)}}$ of $\shifts$ and $\hat{\sigma}$ of $\sigma$. We use the scaled lasso \citep{sun2012scaledlasso} with the same negativity constraint as in~\eqref{eq:plasso}: 
\begin{equation}
\label{eq:scaledlasso}
\left(\hat{\shifts}^{\text{(init)}}, \hat{\sigma}\right) = \argmin_{\shifts \in \shiftset, \sigma > 0}  \frac{\|y - X \shifts\|_2^2}{2\sigma m} + \frac{\sigma}{2} + \lambda_{scaled} \|\shifts\|_1 .
\end{equation}

Problem~\eqref{eq:scaledlasso} can be solved efficiently by iterating between updates of (i) $\hat{\sigma}$ using the closed-form expression $\hat{\sigma} = \|y - X \hat{\shifts}\|_2 / \sqrt{m}$ and (ii) of $\hat{\shifts}$ by solving the constrained lasso problem \eqref{eq:lasso} with tuning parameter $\lambda_{scaled}=\lambda m \hat{\sigma}$. Debiasing is achieved by the corrected update:
\begin{equation}
\label{eq:correction}
\hat{\shifts}_j = \hat{\shifts}_j^{\text{(init)}} + \frac{\langle s_j,y-X\hat{\shifts}^{(\text{init})}\rangle}{\langle s_j,x_j\rangle}.
\end{equation}
where the $s_j$ form a score-system (SS). Intuitively, $s_j$ should form a relaxed orthogonalization of $x_j$ against other column-vectors of $X$. The $s_j$ are used to decorrelate the estimators. We used the strategy of \cite{zhang2014confidence} and take the residuals of a lasso regression of $x_j$ against $X_{-j}$. We also considered the alternative debiasing strategy of \cite{javanmard2013confidence, javanmard2014confidence}, which is based on a pseudo-inverse of $\hat{\Sigma} = \frac{X^TX}{m}$. Their debiased estimate is again a simple update of the initial scaled lasso estimator:
$$\hat{\shifts} = \hat{\shifts}^{(\text{init})} + \frac{1}{m}SX^T \left(y-X\hat{\shifts}^{(\text{init})}\right)$$ 
but the decorrelation matrix $S$ is computed in a so-called colwise inverse approach (CI), by inverting $\hat{\Sigma}$ in a columnwise fashion. Column $s_j$ is solution of the optimization problem :
\begin{equation}
\label{eq:colwise}
\left\{
  \begin{aligned}
    s_j & = \text{argmin}_{s \in \mathbb{R}^{n}} \ s^T\hat{\Sigma}s \\
    &\text{s.t.}\  \|\hat{\Sigma}s - e_j\|_{\infty} \leq \gamma.
  \end{aligned}
\right.
\end{equation}
where $e_j$ is the $j^\text{th}$ canonical vector and $\gamma \geq 0$ is a slack hyperparameter. If $\gamma$ is too small, the problem is not feasible (unless $\hat{\Sigma}$ is non singular). If $\gamma$ is too large, the unique solution is $s_j = 0$.

\paragraph*{Confidence Interval. } \cite{zhang2014confidence} showed that asymptotically $\hat{\shifts} \sim \mathcal{N}\left(\shifts, V\right)$ with the covariance matrix $V$ defined by
\begin{equation}
\label{eq:covar}
v_{ij} = \hat{\sigma}^2 \frac{\langle s_i,s_j\rangle}{\langle s_i,x_i\rangle\langle s_j,x_j\rangle} .
\end{equation}
Similarly, the columnwise-inverse estimator of \cite{javanmard2013confidence} has asymptotic distribution $\mathcal{N}\left(\shifts, V\right)$ with variance matrix $V = S \hat{\Sigma} S^T / m$. For both procedures, the bilateral confidence interval at level $\alpha$ for $\hat{\shifts}_j$ is 
\[
IC_{\alpha}(\hat{\shifts}_j) = \left[ \hat{\shifts}_j \pm \phi^{-1}\left(1-\frac{\alpha}{2}\right) \sqrt{v_{jj}} \right] .
\]
Note that the estimator of the $i^{\text{th}}$ component of $\mu$ can be written $\hat{\mu}_i = t_{i.}^T\hat{\shifts}$ with $t_{i.}^T$ the $i^{\text{th}}$ row of $T$. Its unilateral confidence intervals at level $\alpha$ is thus given by $\left[-\infty, \hat{\mu}_i + \sqrt{t_{i.}^T V t_{i.}} \phi^{-1}\left(1-\alpha\right)\right]$. We can thus simply check whether $0$ falls in the interval to test $\mathcal{H}_{i0} : \{\mu_i = 0\}$ versus $\mathcal{H}_{i1}: \{\mu_i < 0\}$ at level $\alpha$ or compute the p-value of the one-sided test as:

\begin{equation}
\label{eq:sspv}
\pv^\text{ss}_i = \Phi\left(\frac{t_{i.}^T\hat{\shifts}}{\left(t_{i.}^TVt_{i.}\right)^{1/2}}\right).
\end{equation}

\subsubsection{FDR control}

The debiasing procedure achieves marginally consistent interval estimation of the shifts $\shifts$ but additional care is required to control the FDR when testing all components of $\mu$ simultaneously. We use the procedure proposed in \cite{javanmard2019false}, which is specific to debiased lasso estimators, and relies on the $t$-scores $\ts_i = \frac{t_{i.}^T\hat{\shifts}}{\left(t_{i.}^TVt_{i.}\right)^{1/2}}$. Briefly, for FDR control at a given level $\alpha$, let $t_{\text{max}} = \sqrt{2 \log m - 2 \log \log m}$ and set:
\begin{equation*}
t^{\star} = \inf \left\{ 0 \leq t \leq t_{\max} : \frac{2m(1 - \Phi(t))}{R(t) \vee 1} \leq \alpha \right\}
\end{equation*}
where $R(t) = \sum_{i = 1}^m 1_{\{t_i \leq -t\}}$ is the total number of rejections at threshold $t$, or $t^{\star} = \sqrt{2 \log m}$ if the previous expression is empty. Applying the procedure from  \citet{javanmard2019false} strictly would replace $2m$ with $m$ in the numerator, as we're considering one-sided tests instead of two-sided ones for $\mu_i$. However, numerical analysis showed that the extra 2 led to better control of the FDR and we thus kept it. Hypothesis $\mathcal{H}_{i0}$ is rejected if $\ts_i \leq -t^{\star}$ or in term of $q$-values if 
\begin{equation}
\label{eq:ssqv}
\qv^{\text{ss}}_i \coloneqq \frac{\pv^{\text{ss}}_i \alpha}{\Phi(-t^{\star})} \leq \alpha.
\end{equation}
Since $\ts$ itself depends on $\alpha$, the corrected p-values depend on $\alpha$, unlike in the standard BH procedure, where they only depend on the order statistics.

\subsubsection{Algorithm}
\label{sec:algorithms}

The algorithm \ref{algo_hfdr} summarises our procedure. We call it \textit{zazou} for "\textbf{z}-scores \textbf{az} \textbf{O}rnstein-\textbf{U}hlenbeck".

\begin{algorithm}
\caption{Zazou procedure}
\label{algo_hfdr} 
\begin{algorithmic}[1]
\STATE Compute the vector $\pv$ of raw p-values
\STATE Transform it to the vector $z$ of raw z-scores
\FOR{values of $\alpha$ and $\lambda$ varying in a grid}
\STATE Compute $\Sigma$, $R$, $y$ and $X$
\STATE Compute $\hat{\shifts}_{\alpha, \lambda}$ and $\hat{\sigma}_{\alpha, \lambda}$ by solving  \eqref{eq:scaledlasso}
\STATE Compute the BIC criterion~\eqref{eq:bic}
\ENDFOR
\STATE Select parameter values $\hat{\alpha}$ and $\hat{\lambda}$ that minimize the BIC
\STATE Set $\hat{\shifts}^{\text{(init)}} = \hat{\shifts}_{\hat{\alpha}, \hat{\lambda}}$
\STATE Update $\hat{\shifts}^{\text{(init)}}$ according to \eqref{eq:correction} to debias it
\STATE Compute its covariance matrix $\hat{V}$ with \eqref{eq:covar}
\STATE Compute the vector $p$-values $\pv^\text{ss}$ of corrected with \eqref{eq:sspv}
\RETURN Vector of corrected $q$-values $\qv^\text{ss}$ computed from \eqref{eq:ssqv} for a target FDR level $\alpha$. 
\end{algorithmic}
\end{algorithm}
 
\section{Sign-constrained lasso}
\label{sec:Optimization}

Our inference procedure is based on very standard estimates but requires to solve the following constrained lasso problem:
\begin{equation*}
\hat{\shifts} = \argmin_{\shifts \; \text{s.t.} \; T\shifts \in \RR_-^m} \frac{1}{2} \left\|y - X\shifts\right\|_2^2 + \lambda \|\shifts\|_1 .
\end{equation*}
For arbitrary vector $y$ and matrices $X$ and $T$. This a convex problem as both the objective function and feasibility set are convex. We therefore adapt the shooting algorithm \citep{fu1998penalized}, an iterative algorithm used to solve the standard lasso by looping over coordinates and solving simpler unidimensional problem, to our constrained problem.

Let $X_{-j}$ (resp. $\shifts_{-j}$) be the matrix $X$ (resp. vector $\shifts$) deprived of its $j^\text{th}$ column (resp. $j^\text{th}$ coordinate). We can isolate $\shifts_j$ in \eqref{eq:lasso} and decompose the objective as 
$\|y - X\shifts\|^2_2 + \lambda |\shifts| = \| y - z_j - x_j \shifts_j \|^2_2 + \lambda |\shifts_j| + \lambda \|\shifts_{-j}\|_1$ where 
$z_j = X_{-j}\shifts_{-j} \in \RR^{m}$. We can likewise decompose $T\shifts = u_j + v_j\shifts_j$ where $u_j = T_{-j}\shifts_{-j}\in \RR^{m}$ and $v_j = t_j$. When updating $\shifts_j$, we can thus consider the simpler univariate problem in $\theta$:
\begin{equation}
\label{eq:univariate}
\left\{
  \begin{aligned}
    \argmin_{\theta \in \RR} h(\theta) &= \frac{1}{2} \|y - z - x\theta\|^2_2 + \lambda |\theta| \\
    \text{s.t.}\ &u + v\theta \leq 0.
  \end{aligned}
\right.
\end{equation}

Let $I_+ = \{i: v_i > 0\}$ and $I_- = \{i: v_i < 0\}$ and denote 
$\theta_{\max} = \min_{I_{+}} \{ {-u_i}/{v_i} \}$ and 
$\theta_{\min} = \max_{I_{-}} \{ {-u_i}/{v_i} \}$ with the usual conventions that $\max(\emptyset) = -\infty$ and $\min(\emptyset) = +\infty$. Problem~\eqref{eq:univariate} is feasible only if 
(i) $\theta_{\min} \leq \theta_{\max}$ and (ii) for all $i$, $v_i = 0 \Rightarrow u_i \leq 0$, in which case the feasible region is $\lbrack \theta_{\min}, \theta_{\max} \rbrack$. Computing the subgradient $\partial h(\theta)$ of $h$ and looking for values $\theta$ such that $0 \in \partial h(\theta)$ leads to the usual shrinked estimates:

$$
\begin{cases} 
\frac{(y-z)^Tx+\lambda}{x^Tx} & \text{if } (y-z)^Tx < -\lambda, \\ 
\frac{(y-z)^Tx-\lambda}{x^Tx} & \text{if } (y-z)^Tx > \lambda, \\ 
0                             & \text{if } | (y-z)^Tx | < \lambda.
\end{cases}
$$
By convexity of $h$, the solution of \eqref{eq:univariate} can be found by projecting the previous unconstrained minimum to the feasibility set. If problem \eqref{eq:univariate} is feasible, its solution is thus given by 
$$
\theta^{\star} = \begin{cases} 
P_\mathcal{I} \left(\frac{(y-z)^Tx+\lambda}{x^Tx}\right) & \text{if } (y-z)^Tx < -\lambda, \\ 
P_\mathcal{I}\left(\frac{(y-z)^Tx-\lambda}{x^Tx}\right) & \text{if } (y-z)^Tx > \lambda, \\ 
P_\mathcal{I} (0)                             & \text{if } | (y-z)^Tx | < \lambda,
\end{cases}
$$

where $P_\mathcal{I} : u \mapsto \max(\theta_{\min}, \min(u, \theta_{\max}))$ is the projection of $u$ on the segment $\mathcal{I} = \lbrack \theta_{\min}, \theta_{\max} \rbrack$.

\section{Synthetic Data \label{sec:Simul}}

\subsection{Metagenomics}

Metagenomics data are made up of three components. The first component is the count or abundance matrix $X = (x_{ij})$, with $1 \leq i \leq m$ and $1 \leq j \leq p$, which represents the quantity of taxa $i$ in sample $j$. The second component is a set of sample covariates, such as disease status, environmental conditions, group, etc. The final component is a phylogenetic tree  which captures the shared evolutionary history of all taxa. When performing DAT, we are interested in taxa whose abundance is significantly associated to a covariate. 

Most DAT procedures proceed with univariate tests (one test per species) followed by a correction procedure. In the synthetic datasets, we consider discrete covariates only. Dozens of full-fledged testing pipelines are published each year, including some designed with omics data in mind. Since our goal is this study is to compare correction procedures rather than full testing procedures, we use Wilcoxon or Kruskall-Wallis tests, which are classical and widespread non parametric tests in metagenomics.  

\subsection{Simulations}

\paragraph*{Simulation scheme. } We use the following simulation scheme:
\begin{enumerate}
\item start with a homogeneous dataset,
\item assign each sample to group A or B at random
\item select differentially abundant taxa in a phylogenetically consistent manner (diffentially abundant taxa)
\item apply a fold-change to the observed abundance of diffentially abundant taxa in group B. 
\end{enumerate}

This non-parametric simulation scheme was previously used in \cite{bichat2020incorporating}. We considered two variants for step 3, respectively called \emph{positive} and \emph{negative}. In the negative variant, differentially abundant taxa were selected randomly across the tree, so that the phylogeny is not informative. In the positive variant, taxa are instead selected in a phylogenetically consistent manner. Formally, the phylogeny was first used to compute the cophenetic \citep{sneath1973numerical} distance matrix between taxa. A partioning around medoids algorithm was then used to create cluster of related species. One or more clusters were then picked at random and all species in those clusters were selected as differentially abundant. 

For each fold-change ($\text{fc} \in \{3, 5, 10\}$), 500 simulated datasets were created, with a proportion of differentially abundant species ranging from 3~\% to 35~\%. For each simulation, we corrected $p$-values using no correction (Raw), BH procedure (BH), BY procedure (BY), \texttt{StructFDR} (TF) or our procedure with either score system (SS) or colwise inverse debiasing (CI), targeting in all instances a 5\% FDR level. We compared the 6 procedures in terms of True Positive Rate (TPR), nominal FDR and AUC (Area Under the Curve). 

\paragraph*{Positive simulations.}

\begin{figure}[h!]
\begin{center}
\includegraphics[width=\linewidth]{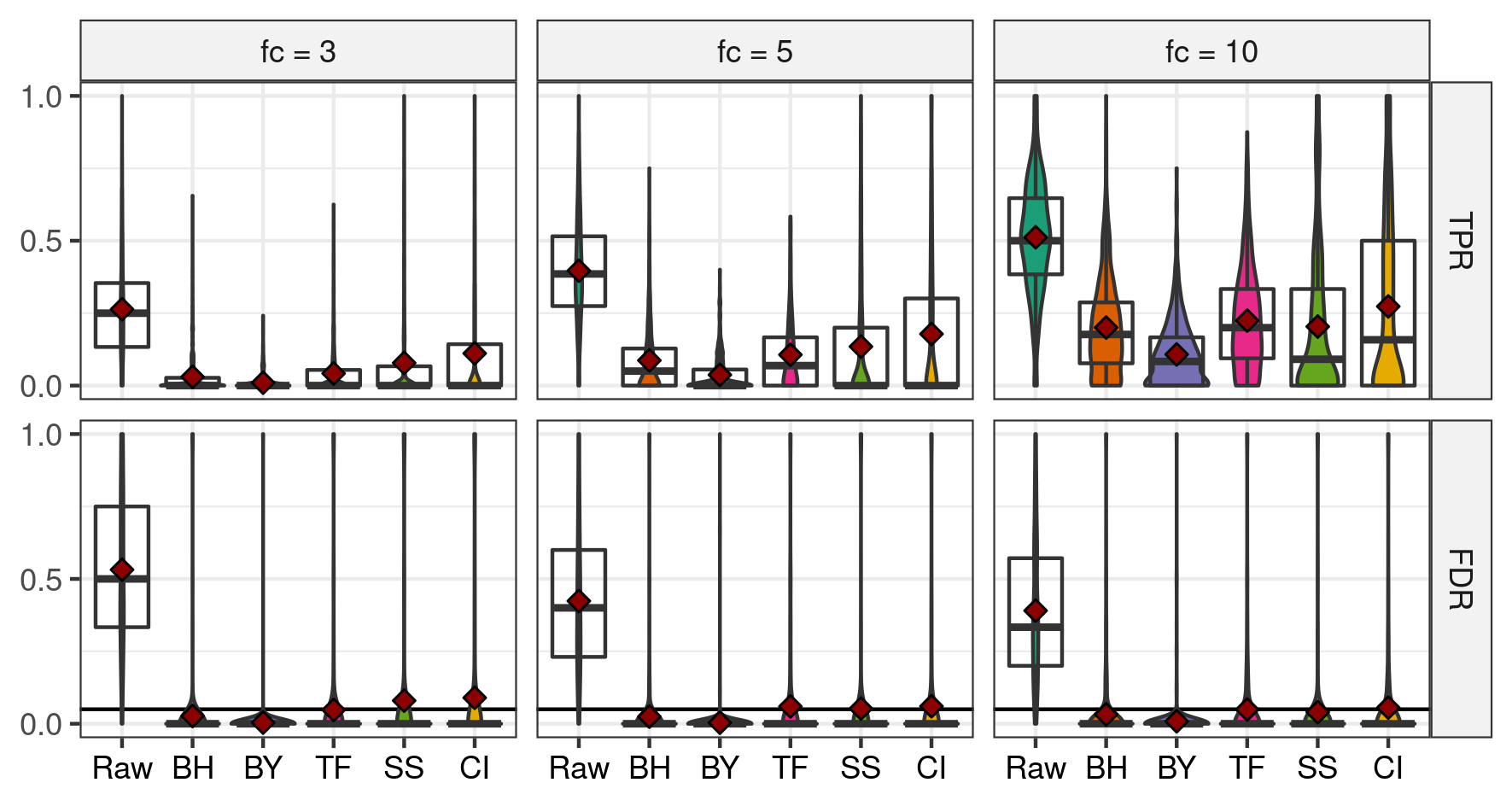}
\end{center}
\caption{\label{fig:tprfdr} Boxplots and average (red point) TPR and FDR across positive simulation settings. Each facet corresponds to a different fold-change (fc) and each boxplot is computed over 500 simulation replicates. All corrections control the FDR at the target level or slightly above but \texttt{zazou} (SS and CI) achieve higher TPR, especially for small fold changes.}
\end{figure}

The results of positive simulations (\emph{i.e.} where the phylogeny is informative) are shown in Figure~\ref{fig:tprfdr}. All correction methods have controlled the FDR at the target rate or below when the fold change is larger than 5. For smaller fold changes, both SS and CI variations of \texttt{zazou} exhibit nominal FDR slightly above the target level (up to 9\% in the worst case). In all settings, BY had the lowest TPR, whereas TF was comparable to vanilla BH, in line with results of \citet{bichat2020incorporating}. Finally, \texttt{zazou} (both SS and CI variations) had the best overall TPR, with largest gains observed in the lowest fold-change setting. 

\begin{figure}[h!]
\begin{center}
\includegraphics[width=\linewidth]{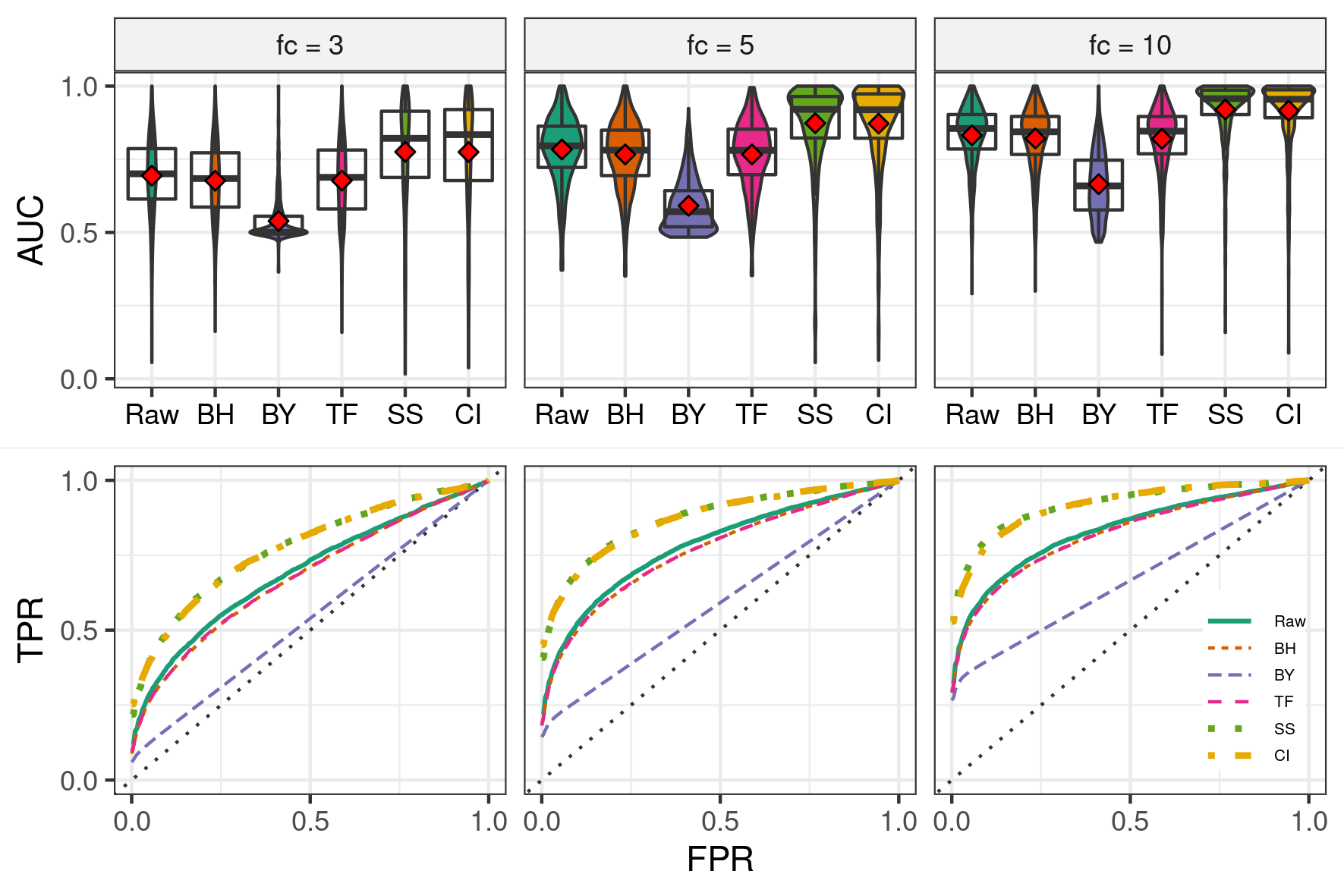}
\end{center}
\caption{\label{fig:aucroc} AUC boxplots (top) and average ROC curves (bottom) across positive simulations settings. Facets correspond to fold-changes (fc). ROC curves are computed for each simulation and linearly interpolated over a fixed grid before being averaged. Each boxplot and each curve are computed over 500 replicates. In all settings, SS/CI have the highest AUC / ROC curve, followed by BH/TF while BY has the lowest values.}
\end{figure}

The higher than intended FDR of \texttt{zazou} methods suggests that the problem of finding an adequate threshold for $\pv_i^{ss}$ is not completely solved by \citet{javanmard2019false} procedure. To assess the performance of \texttt{zazou} in a threshold-independent manner, we also compared the AUC of all procedures. Fig.~\ref{fig:aucroc} shows that \texttt{zazou} (both variants) has higher AUC than all other methods. As reported previously, TF and BH are at the same level and BY has the lowest ROC curve. Focus on the beginning of left hand side side of the curve shows that \texttt{zazou} is more efficient starting from the first discoveries. 

\paragraph*{Negative simulations.}

\begin{figure}[h!]
\begin{center}
\includegraphics[width=0.5\linewidth]{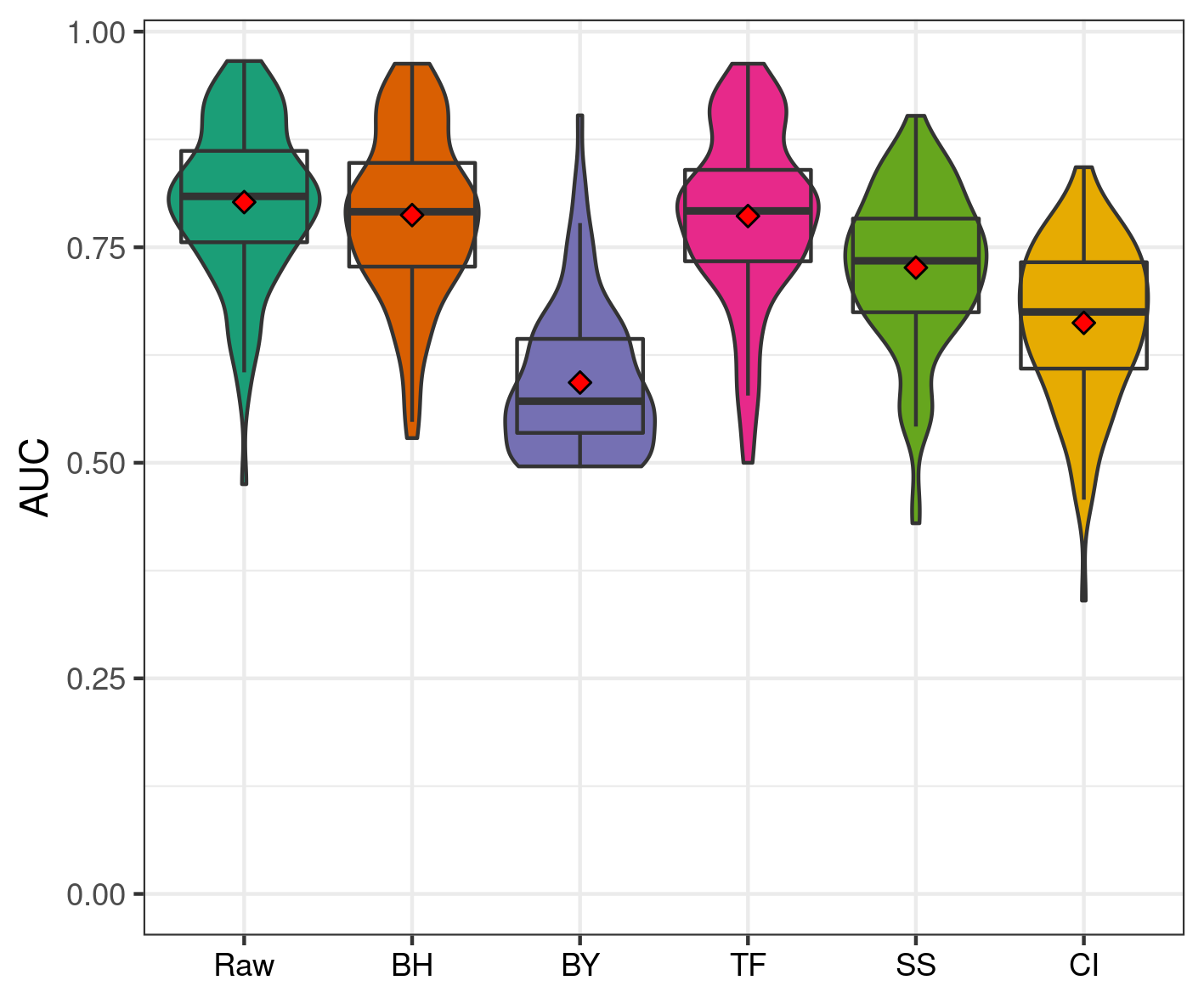}
\end{center}
\caption{\label{fig:aucneg} AUC boxplots (computed over 500 replicates) in negative simulations. BH outperforms SS and CI, highlighting the cost of imposing a mispecified hierarchical structure.}
\end{figure}

The negative simulations are designed to assess the robustness of our algorithm with respect to uninformative  phylogenies, or equivalently mispecified hierarchies. Fig.~\ref{fig:aucneg} shows that, as expected, standard BH outperforms competing methods (in terms of AUC) when the tree is mispecified. Forcing an inadequate tree structure results in AUC losses ranging from 15 to 20 percentage points compared to no structure. The puzzling lack of AUC loss for the TF procedure is explained by an implementation trick: TF always performs BH correction in parallel to its hierarchical procedure and falls back to BH when the hierarchical procedure detects much fewer species than BH \citep{bichat2020incorporating, xiao2017false}.

\section{Application \label{sec:Appli}}

We use our \texttt{zazou} procedure on a gut microbiota dataset from the Fiji Islands \citep{brito2016mobile, pasolli2017accessible} to identify species that are differentially abundant between adults and children. The data sets consists in the abundances of $p=387$ species among $n = 146$ islanders, split into 112 adults and 34 children. 

To mimick the simulation study, we used Wilcoxon tests for the univariate tests. Without correction, 21 species were detected as differentially abundant at the 5\% level. None of them remained significant after correction by BH, BY, TreeFDR or treeclimbR. By contrast, \texttt{zazou} detected differentially abundant species with both desparsification methods: 17 for SS and 6 for CI. 

\begin{figure}[h!]
\begin{center}
\includegraphics[width=\linewidth]{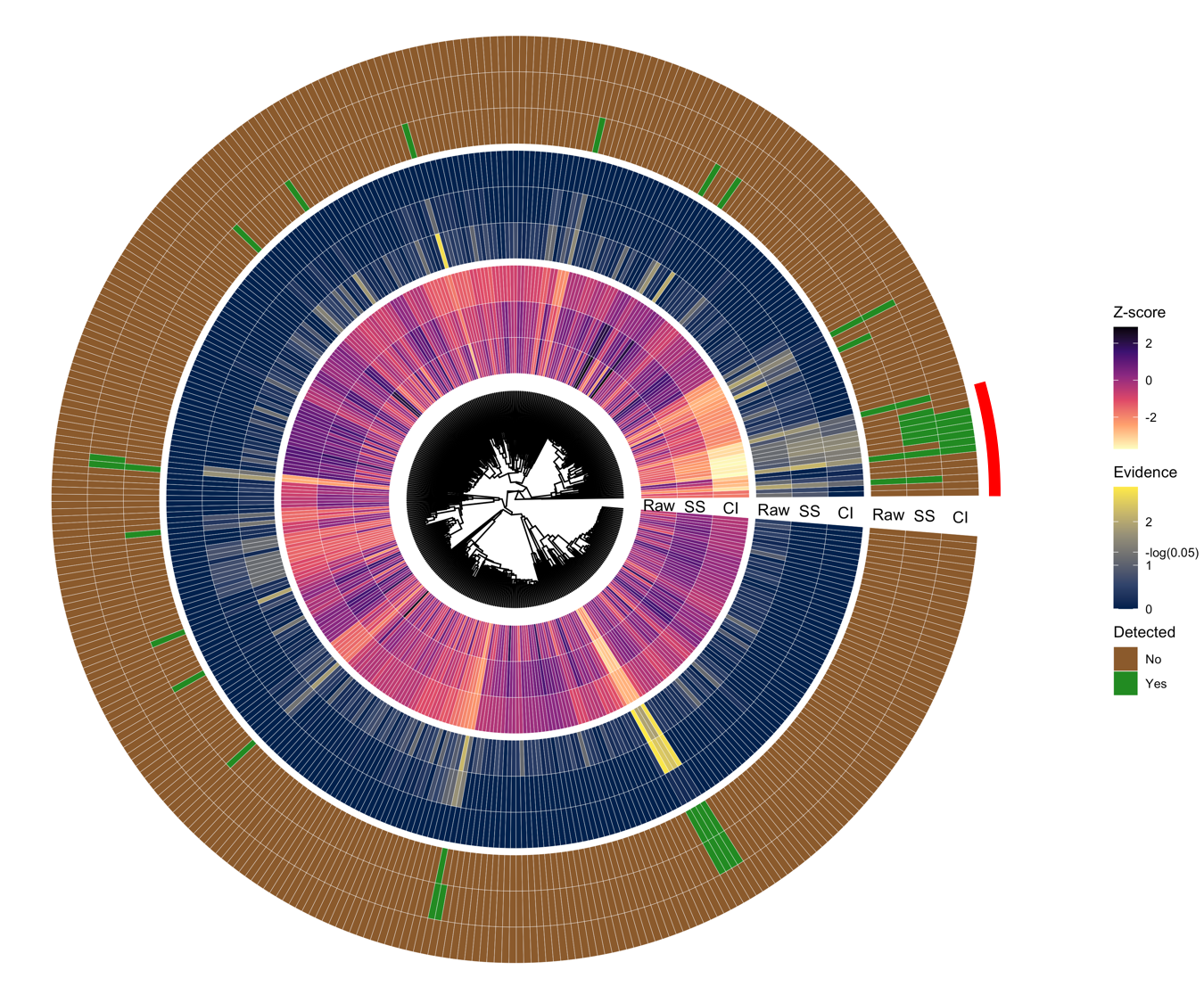}
\end{center}
\caption{\label{fig:heattree} Phylogeny of the 387 species from the Fidji dataset with associated $z$-scores (inner circle), evidence (middle circle) and detection status (outer circle) under different correction procedures. Species detected by \texttt{zazou} are generally close-by on the tree and often, but not always, detected by raw $p$-values. The red strip highlight the smoothing property of the procedure in a subtree where individual species are not detected when using independant univariate tests but are detected when accounting for the hierarchical structure.}
\end{figure}

Fig.~\ref{fig:heattree} shows that they are not a strict subset of the 21 detected with no correction. Smoothing salvages some species that are closely related to one of the 21 without being significant on their own (red box in the figure). It also illustrate some numerical problems associated with colwise-inverse debiasing, which is highly sensitive  to the choice of the slack hyperparameter  $\gamma$. The window of relevant values for $\gamma$ is narrow and too large or too small values $\gamma$ respectively lead to no correction or a faulty p-value correction. 

\section{Conclusion}

In this work, we introduced \texttt{zazou}, a new method for correcting $p$ values in a hierarchical context. \texttt{zazou} is based on recasting the testing problem as a regression problem, under the framework of stochastic processes on an ultrametric tree, and using the tree topology as a regularization parameter. 

It outperforms competing methods, hierarchical (TreeFDR, TreeclimbR) or not (BH, BY) in terms of AUC but this does not translate immediately to superior results in terms of FDR and TPR. The threshold for rejecting hypotheses is turned out to be quite difficult to calibrate while controling the FDR and warrants further work. 

There are several other parts of the procedure that are not as powerful as expected. First, the BIC step used to select $\lambda$ and in turn the number of shifts tends to choose models with very few shifts, and sometimes even none. In such instances, the relevance of the debiasing step is limited.  Second, the correction procedure proposed by \cite{javanmard2019false} is too conservative for our purpose. It was indeed developed to control both the FDR and the directional FDR (\emph{i.e.} proportion of Type S errors, where the effect size have the wrong sign, in the discoveries) whereas we only need to control the former. For both these steps, specific developments taking into account the sign constraint on $\hat{\mu}$ and the structure of the topology matrix of tree $\tree$ could lead to better performances for \texttt{zazou}.


\begin{thebibliography}{48}
\providecommand{\natexlab}[1]{#1}
\providecommand{\url}[1]{\texttt{#1}}
\expandafter\ifx\csname urlstyle\endcsname\relax
  \providecommand{\doi}[1]{doi: #1}\else
  \providecommand{\doi}{doi: \begingroup \urlstyle{rm}\Url}\fi

\bibitem[Ambroise et~al.(2019)Ambroise, Dehman, Neuvial, Rigaill, and
  Vialaneix]{ambroise2019adjacency}
Christophe Ambroise, Alia Dehman, Pierre Neuvial, Guillem Rigaill, and Nathalie
  Vialaneix.
\newblock Adjacency-constrained hierarchical clustering of a band similarity
  matrix with application to genomics.
\newblock \emph{Algorithms for Molecular Biology}, 14\penalty0 (1):\penalty0
  22, 2019.

\bibitem[Bastide et~al.(2017)Bastide, Mariadassou, and
  Robin]{bastide2017detection}
Paul Bastide, Mahendra Mariadassou, and St{\'e}phane Robin.
\newblock Detection of adaptive shifts on phylogenies by using shifted
  stochastic processes on a tree.
\newblock \emph{Journal of the Royal Statistical Society: Series B (Statistical
  Methodology)}, 79\penalty0 (4):\penalty0 1067--1093, 2017.

\bibitem[Benjamini and Heller(2007)]{benjamini2007false}
Yoav Benjamini and Ruth Heller.
\newblock False discovery rates for spatial signals.
\newblock \emph{Journal of the American Statistical Association}, 102\penalty0
  (480):\penalty0 1272--1281, 2007.

\bibitem[Benjamini and Hochberg(1995)]{benjamini1995controlling}
Yoav Benjamini and Yosef Hochberg.
\newblock Controlling the false discovery rate: a practical and powerful
  approach to multiple testing.
\newblock \emph{Journal of the Royal statistical society: series B
  (Methodological)}, 57\penalty0 (1):\penalty0 289--300, 1995.

\bibitem[Benjamini and Yekutieli(2001)]{benjamini2001control}
Yoav Benjamini and Daniel Yekutieli.
\newblock The control of the false discovery rate in multiple testing under
  dependency.
\newblock \emph{Annals of statistics}, pages 1165--1188, 2001.

\bibitem[Bichat et~al.(2020)Bichat, Plassais, Ambroise, and
  Mariadassou]{bichat2020incorporating}
Antoine Bichat, Jonathan Plassais, Christophe Ambroise, and Mahendra
  Mariadassou.
\newblock Incorporating phylogenetic information in microbiome differential
  abundance studies has no effect on detection power and fdr control.
\newblock \emph{Frontiers in Microbiology}, 11:\penalty0 649, 2020.
\newblock ISSN 1664-302X.
\newblock \doi{10.3389/fmicb.2020.00649}.
\newblock URL
  \url{https://www.frontiersin.org/article/10.3389/fmicb.2020.00649}.

\bibitem[Blanchard et~al.(2020)Blanchard, Neuvial, and
  Roquain]{blanchard2020dependency}
Gilles Blanchard, Pierre Neuvial, and Etienne Roquain.
\newblock {Post hoc confidence bounds on false positives using reference
  families}.
\newblock \emph{The Annals of Statistics}, 48\penalty0 (3):\penalty0 1281 --
  1303, 2020.
\newblock \doi{10.1214/19-AOS1847}.
\newblock URL \url{https://doi.org/10.1214/19-AOS1847}.

\bibitem[Bland and Altman(1995)]{bland1995multiple}
J~Martin Bland and Douglas~G Altman.
\newblock Multiple significance tests: the bonferroni method.
\newblock \emph{Bmj}, 310\penalty0 (6973):\penalty0 170, 1995.

\bibitem[Brito et~al.(2016)Brito, Yilmaz, Huang, Xu, Jupiter, Jenkins,
  Naisilisili, Tamminen, Smillie, Wortman, et~al.]{brito2016mobile}
Ilana~L Brito, S~Yilmaz, K~Huang, Liyi Xu, Stacy~D Jupiter, Aaron~P Jenkins,
  Waisea Naisilisili, M~Tamminen, CS~Smillie, Jennifer~R Wortman, et~al.
\newblock Mobile genes in the human microbiome are structured from global to
  individual scales.
\newblock \emph{Nature}, 535\penalty0 (7612):\penalty0 435--439, 2016.

\bibitem[Bush and Moore(2012)]{bush2012genome}
William~S Bush and Jason~H Moore.
\newblock Genome-wide association studies.
\newblock \emph{PLoS Comput Biol}, 8\penalty0 (12):\penalty0 e1002822, 2012.

\bibitem[Chen(2018)]{structfdr2018}
Jun Chen.
\newblock \emph{StructFDR: False Discovery Control Procedure Integrating the
  Prior Structure Information}, 2018.
\newblock URL \url{https://CRAN.R-project.org/package=StructFDR}.
\newblock R package version 1.3.

\bibitem[Cremers et~al.(2017)Cremers, Wager, and Yarkoni]{cremers2017relation}
Henk~R Cremers, Tor~D Wager, and Tal Yarkoni.
\newblock The relation between statistical power and inference in fmri.
\newblock \emph{PloS one}, 12\penalty0 (11):\penalty0 e0184923, 2017.

\bibitem[Dunn and Gipson(1977)]{Dunn1977}
James~E Dunn and Phillip~S Gipson.
\newblock Analysis of radio telemetry data in studies of home range.
\newblock \emph{Biometrics}, pages 85--101, 1977.

\bibitem[Eickhoff et~al.(2015)Eickhoff, Thirion, Varoquaux, and
  Bzdok]{eickhoff2015connectivity}
Simon~B Eickhoff, Bertrand Thirion, Ga{\"e}l Varoquaux, and Danilo Bzdok.
\newblock Connectivity-based parcellation: Critique and implications.
\newblock \emph{Human brain mapping}, 36\penalty0 (12):\penalty0 4771--4792,
  2015.

\bibitem[Fan and Tang(2013)]{fan2013}
Yingying Fan and Cheng~Yong Tang.
\newblock Tuning parameter selection in high dimensional penalized likelihood.
\newblock \emph{Journal of the Royal Statistical Society. Series B (Statistical
  Methodology)}, 75\penalty0 (3):\penalty0 531--552, 2013.
\newblock ISSN 13697412, 14679868.
\newblock URL \url{http://www.jstor.org/stable/24772736}.

\bibitem[Freckleton et~al.(2003)Freckleton, Harvey, and Pagel]{Freckleton2003}
Robert~P. Freckleton, Paul~H. Harvey, and Mark Pagel.
\newblock Bergmann's rule and body size in mammals.
\newblock \emph{The American Naturalist}, 161\penalty0 (5):\penalty0 821--825,
  May 2003.
\newblock \doi{10.1086/374346}.
\newblock URL \url{https://doi.org/10.1086/374346}.

\bibitem[Fu(1998)]{fu1998penalized}
Wenjiang~J Fu.
\newblock Penalized regressions: the bridge versus the lasso.
\newblock \emph{Journal of computational and graphical statistics}, 7\penalty0
  (3):\penalty0 397--416, 1998.

\bibitem[Goeman and Finos(2012)]{goeman2012inheritance}
Jelle~J Goeman and Livio Finos.
\newblock The inheritance procedure: multiple testing of tree-structured
  hypotheses.
\newblock \emph{Statistical applications in genetics and molecular biology},
  11\penalty0 (1):\penalty0 1--18, 2012.

\bibitem[Huang et~al.(2021)Huang, Soneson, Germain, Schmidt, Von~Mering, and
  Robinson]{huang2021treeclimbr}
Ruizhu Huang, Charlotte Soneson, Pierre-Luc Germain, Thomas~SB Schmidt,
  Christian Von~Mering, and Mark~D Robinson.
\newblock treeclimbr pinpoints the data-dependent resolution of hierarchical
  hypotheses.
\newblock \emph{Genome biology}, 22\penalty0 (1):\penalty0 1--21, 2021.

\bibitem[Javanmard and Montanari(2013)]{javanmard2013confidence}
Adel Javanmard and Andrea Montanari.
\newblock Confidence intervals and hypothesis testing for high-dimensional
  statistical models.
\newblock In \emph{Advances in Neural Information Processing Systems}, pages
  1187--1195, 2013.

\bibitem[Javanmard and Montanari(2014)]{javanmard2014confidence}
Adel Javanmard and Andrea Montanari.
\newblock Confidence intervals and hypothesis testing for high-dimensional
  regression.
\newblock \emph{The Journal of Machine Learning Research}, 15\penalty0
  (1):\penalty0 2869--2909, 2014.

\bibitem[Javanmard et~al.(2019)Javanmard, Javadi, et~al.]{javanmard2019false}
Adel Javanmard, Hamid Javadi, et~al.
\newblock False discovery rate control via debiased lasso.
\newblock \emph{Electronic Journal of Statistics}, 13\penalty0 (1):\penalty0
  1212--1253, 2019.

\bibitem[Khabbazian et~al.(2016)Khabbazian, Kriebel, Rohe, and
  An{\'e}]{khabbazian2016fast}
Mohammad Khabbazian, Ricardo Kriebel, Karl Rohe, and C{\'e}cile An{\'e}.
\newblock Fast and accurate detection of evolutionary shifts in
  ornstein--uhlenbeck models.
\newblock \emph{Methods in Ecology and Evolution}, 7\penalty0 (7):\penalty0
  811--824, 2016.

\bibitem[Kim et~al.(2010)Kim, Roquain, and van~de Wiel]{kim2010spatial}
Kyung~In Kim, Etienne Roquain, and Mark~A van~de Wiel.
\newblock Spatial clustering of array cgh features in combination with
  hierarchical multiple testing.
\newblock \emph{Statistical applications in genetics and molecular biology},
  9\penalty0 (1), 2010.

\bibitem[Lande(1976)]{Lande1976}
Russell Lande.
\newblock Natural {S}election and {R}andom {G}enetic {D}rift in {P}henotypic
  {E}volution.
\newblock \emph{Evolution}, 30\penalty0 (2):\penalty0 314--334, June 1976.
\newblock \doi{10.1111/j.1558-5646.1976.tb00911.x}.
\newblock URL \url{https://doi.org/10.1111/j.1558-5646.1976.tb00911.x}.

\bibitem[MacLean et~al.(2021)MacLean, Lytras, Weaver, Singer, Boni, Lemey,
  Pond, and Robertson]{maclean2021natural}
Oscar~A MacLean, Spyros Lytras, Steven Weaver, Joshua~B Singer, Maciej~F Boni,
  Philippe Lemey, Sergei L~Kosakovsky Pond, and David~L Robertson.
\newblock Natural selection in the evolution of sars-cov-2 in bats created a
  generalist virus and highly capable human pathogen.
\newblock \emph{PLoS biology}, 19\penalty0 (3):\penalty0 e3001115, 2021.

\bibitem[Mann and Whitney(1947)]{mann1947test}
Henry~B Mann and Donald~R Whitney.
\newblock On a test of whether one of two random variables is stochastically
  larger than the other.
\newblock \emph{The annals of mathematical statistics}, pages 50--60, 1947.

\bibitem[Matsen~IV and Evans(2013)]{matsen2013edgepca}
Frederick~A. Matsen~IV and Steven~N. Evans.
\newblock Edge principal components and squash clustering: Using the special
  structure of phylogenetic placement data for sample comparison.
\newblock \emph{PLOS ONE}, 8\penalty0 (3):\penalty0 1--15, 03 2013.
\newblock \doi{10.1371/journal.pone.0056859}.
\newblock URL \url{https://doi.org/10.1371/journal.pone.0056859}.

\bibitem[McLachlan and Peel(2000)]{mclachlan2000finite}
G~McLachlan and D~Peel.
\newblock Finite mixture models.,(john wiley \& sons: New york.).
\newblock 2000.

\bibitem[McLachlan et~al.(2005)McLachlan, Do, and
  Ambroise]{mclachlan2005analyzing}
Geoffrey~J McLachlan, Kim-Anh Do, and Christophe Ambroise.
\newblock \emph{Analyzing microarray gene expression data}, volume 422.
\newblock John Wiley \& Sons, 2005.

\bibitem[Meinshausen(2008)]{meinshausen2008hierarchical}
Nicolai Meinshausen.
\newblock Hierarchical testing of variable importance.
\newblock \emph{Biometrika}, 95\penalty0 (2):\penalty0 265--278, 2008.

\bibitem[Nåsell(1999)]{Nasell1999}
I.~Nåsell.
\newblock On the time to extinction in recurrent epidemics.
\newblock \emph{Journal of the Royal Statistical Society: Series B (Statistical
  Methodology)}, 61\penalty0 (2):\penalty0 309--330, 1999.
\newblock \doi{https://doi.org/10.1111/1467-9868.00178}.
\newblock URL
  \url{https://rss.onlinelibrary.wiley.com/doi/abs/10.1111/1467-9868.00178}.

\bibitem[Pasolli et~al.(2017)Pasolli, Schiffer, Manghi, Renson, Obenchain,
  Truong, Beghini, Malik, Ramos, Dowd, et~al.]{pasolli2017accessible}
Edoardo Pasolli, Lucas Schiffer, Paolo Manghi, Audrey Renson, Valerie
  Obenchain, Duy~Tin Truong, Francesco Beghini, Faizan Malik, Marcel Ramos,
  Jennifer~B Dowd, et~al.
\newblock Accessible, curated metagenomic data through experimenthub.
\newblock \emph{Nature methods}, 14\penalty0 (11):\penalty0 1023, 2017.

\bibitem[Renaux et~al.(2020)Renaux, Buzdugan, Kalisch, and
  B{\"u}hlmann]{renaux2020hierarchical}
Claude Renaux, Laura Buzdugan, Markus Kalisch, and Peter B{\"u}hlmann.
\newblock Hierarchical inference for genome-wide association studies: a view on
  methodology with software.
\newblock \emph{Computational Statistics}, 35\penalty0 (1):\penalty0 1--40,
  2020.

\bibitem[Sankaran and Holmes(2014)]{sankaran2014structssi}
Kris Sankaran and Susan Holmes.
\newblock structssi: simultaneous and selective inference for grouped or
  hierarchically structured data.
\newblock \emph{Journal of statistical software}, 59\penalty0 (13):\penalty0 1,
  2014.

\bibitem[Segata et~al.(2011)Segata, Izard, Waldron, Gevers, Miropolsky,
  Garrett, and Huttenhower]{segata2011metagenomic}
Nicola Segata, Jacques Izard, Levi Waldron, Dirk Gevers, Larisa Miropolsky,
  Wendy~S Garrett, and Curtis Huttenhower.
\newblock Metagenomic biomarker discovery and explanation.
\newblock \emph{Genome biology}, 12\penalty0 (6):\penalty0 1--18, 2011.

\bibitem[Sesia et~al.(2020)Sesia, Katsevich, Bates, Cand{\`e}s, and
  Sabatti]{sesia2020multi}
Matteo Sesia, Eugene Katsevich, Stephen Bates, Emmanuel Cand{\`e}s, and Chiara
  Sabatti.
\newblock Multi-resolution localization of causal variants across the genome.
\newblock \emph{Nature communications}, 11\penalty0 (1):\penalty0 1--10, 2020.

\bibitem[Silverman et~al.(2017)Silverman, Washburne, Mukherjee, and
  David]{silverman2017philr}
Justin~D Silverman, Alex~D Washburne, Sayan Mukherjee, and Lawrence~A David.
\newblock A phylogenetic transform enhances analysis of compositional
  microbiota data.
\newblock \emph{{eLife}}, 6, February 2017.
\newblock \doi{10.7554/elife.21887}.
\newblock URL \url{https://doi.org/10.7554/elife.21887}.

\bibitem[Sneath et~al.(1973)Sneath, Sokal, et~al.]{sneath1973numerical}
Peter~HA Sneath, Robert~R Sokal, et~al.
\newblock \emph{Numerical taxonomy. The principles and practice of numerical
  classification.}
\newblock 1973.

\bibitem[Sun and Zhang(2012)]{sun2012scaledlasso}
Tingni Sun and Cun-Hui Zhang.
\newblock {Scaled sparse linear regression}.
\newblock \emph{Biometrika}, 99\penalty0 (4):\penalty0 879--898, 09 2012.
\newblock ISSN 0006-3444.
\newblock \doi{10.1093/biomet/ass043}.
\newblock URL \url{https://doi.org/10.1093/biomet/ass043}.

\bibitem[Tang et~al.(2017)Tang, Chen, Alekseyenko, and Li]{tang2017general}
Zheng-Zheng Tang, Guanhua Chen, Alexander~V Alekseyenko, and Hongzhe Li.
\newblock A general framework for association analysis of microbial communities
  on a taxonomic tree.
\newblock \emph{Bioinformatics}, 33\penalty0 (9):\penalty0 1278--1285, 2017.

\bibitem[Tibshirani(1996)]{tibshirani1996regression}
Robert Tibshirani.
\newblock Regression shrinkage and selection via the lasso.
\newblock \emph{Journal of the Royal Statistical Society: Series B
  (Methodological)}, 58\penalty0 (1):\penalty0 267--288, 1996.

\bibitem[Tukey(1953)]{tukey1953problem}
John~Wilder Tukey.
\newblock The problem of multiple comparisons.
\newblock \emph{Multiple comparisons}, 1953.

\bibitem[Tusher et~al.(2001)Tusher, Tibshirani, and
  Chu]{tusher2001significance}
Virginia~Goss Tusher, Robert Tibshirani, and Gilbert Chu.
\newblock Significance analysis of microarrays applied to the ionizing
  radiation response.
\newblock \emph{Proceedings of the National Academy of Sciences}, 98\penalty0
  (9):\penalty0 5116--5121, 2001.

\bibitem[Wilcoxon(1992)]{wilcoxon1992individual}
Frank Wilcoxon.
\newblock Individual comparisons by ranking methods.
\newblock In \emph{Breakthroughs in statistics}, pages 196--202. Springer,
  1992.

\bibitem[Xiao et~al.(2017)Xiao, Cao, and Chen]{xiao2017false}
Jian Xiao, Hongyuan Cao, and Jun Chen.
\newblock False discovery rate control incorporating phylogenetic tree
  increases detection power in microbiome-wide multiple testing.
\newblock \emph{Bioinformatics}, 33\penalty0 (18):\penalty0 2873--2881, 2017.

\bibitem[Yekutieli(2008)]{yekutieli2008hierarchical}
Daniel Yekutieli.
\newblock Hierarchical false discovery rate--controlling methodology.
\newblock \emph{Journal of the American Statistical Association}, 103\penalty0
  (481):\penalty0 309--316, 2008.

\bibitem[Zhang and Zhang(2014)]{zhang2014confidence}
Cun-Hui Zhang and Stephanie~S Zhang.
\newblock Confidence intervals for low dimensional parameters in high
  dimensional linear models.
\newblock \emph{Journal of the Royal Statistical Society: Series B (Statistical
  Methodology)}, 76\penalty0 (1):\penalty0 217--242, 2014.

\end{thebibliography}
\end{document}